# Statistical Description for Assumption-free Single-shot Measurement of Femtosecond Laser Pulse Parameters via Two-photon-induced Photocurrents

Eric R. Tkaczyk, Sylvain Rivet, Lionel Canioni, Stéphane Santran and Laurent Sarger

*Abstract*—Through examining the product of the mathematical variance of intensity with respect to time and frequency, we arrive at a temporal characterization of laser pulses through parameters for pulse duration, group delay dispersion and temporal form. These statistics, which are sufficient to predict subsequent pulse behavior, are recoverable in a simple experiment, measuring the two-photon-induced photocurrents in three nonlinear diodes. With only two photodiodes, we demonstrate that pulse durations as low as several tens of femtoseconds can be easily measured in a single shot if the usual assumptions of pulse form and dispersion are made as in the more difficult autocorrelation setup.

## I. Introduction

IN current ultrafast optical science, pulse duration and dispersion are some of the most elusive characteristics to measure, yet they are key in anticipating the nonlinear behavior and nonlinear effects created during subsequent wave propagation that are the primary interest of this field [1]. The standard procedure to estimate pulse duration relies on an experimentally tedious autocorrelation measurement [2], [3], [4]. Since a direct linear autocorrelation measurement only gives information about the coherence length of the pulse, the measurement in a linear diode necessitates an SHG signal generated in a thin nonlinear crystal. Now the advent of inexpensive nonlinear diodes has allowed direct second-order intensity measurements, which have been implemented in a slightly less elaborate autocorrelation measurement, thereby avoiding the spectral filtering effect of the nonlinear crystal [5].

To extract pulse duration information from the autocorrelation signal, a specific temporal form, usually gaussian or hyperbolic secant, is assumed for all pulses from the laser. In actuality, the envelope is most likely something between a hyperbolic secant and a gaussian, the former being the solitonic propagation form selected in the laser cavity, and the latter being the form approached after traversing significant quantities of dispersive media. The pulse duration is usually defined as the FWHM of the assumed pulse intensity as a function of time, which may not always reflect the actual pulse duration. Although there exist also methods like frequency-resolved optical gating [6], frequency domain pulse measurement [7], [8], and the SPIDER technique [9], [10] for complete E-field amplitude and phase recovery, such a detailed description of the femtosecond pulse is usually not mandatory and does not justify the required experimental labor. Further, even if they do not presuppose a temporal form, these latter methods still either assume uniform laser pulses during the data set acquisition or require tedious experimental preparation. Moreover, in this ultrashort light pulse, one has to worry about the carrier frequency all along the pulse envelope. In an ideal, so-called chirpless pulse, the instantaneous frequency is stationary.

In light of these considerations, we propose a simple, parametric statistical description of the laser pulse without the maze of unwieldy calculations of a complete E-field recovery.

## II. Statistical Description of Laser Pulses

As widely accepted for the space domain, the proposed description relies upon the variance of pulse intensity $I$ with respect to time, $\Delta t^2$, and frequency, $\Delta \omega^2$, each being defined as an rms value in the normal fashion in terms of moments:

$$\Delta t^2 = \langle t^2 \rangle - \langle t \rangle^2, \qquad (1)$$

where the normalized intensity as a function of time is the probability distribution function:

$$\langle f(t) \rangle = \frac{\int f(t) |E_{(t)}|^2 dt}{\int |E_{(t)}|^2 dt} \quad (2)$$

We adopt $\Delta t = (\Delta t^2)^{1/2}$ as our definition of pulse duration $\tau$, conferring the benefit of an unambiguous measure that is resistant to possible asymmetries and noise in temporal form. A major advantage of our definition over FWHM is that it also reflects satellite peaks that may occur around the primary pulse, and to which the media may respond. This however, does also have the drawback of increased sensitivity towards far away but weak satellite pulses.

The uncertainty principle dictates that the product $\Delta t\, \Delta \omega$ must be at least 1/2. We define a form factor $M^2$ [11] for the pulse corresponding to the area it would occupy in the time-frequency Wigner plane [12] if the pulse had no dispersion:

$$\Delta \omega\ \Delta t_{\varphi=0} = \frac{M^2}{2} \quad (3)$$

$M^2$ is always at least 1, this minimal value being strictly achieved for a Gaussian pulse. $M^2$ for a hyperbolic secant is 1.05, 1.4 for a Lorentzian, and 6.54 for a square pulse. $M^2$ can be used to place the form of the impulse along this scale.

Now we consider the pulse propagating through a dispersive medium. The field $\tilde{E}_{(\omega)}$ in the frequency domain is written as amplitude and phase with weak frequency dependence up to the second order:

$$\tilde{E}_{(\omega)} = \tilde{A}_{(\omega)} e^{i\tilde{\varphi}_{(\omega)}} = FT\{E_{(t)}\} \quad (4)$$

$$\tilde{\varphi}_{(\omega)} \approx \dot{\tilde{\varphi}}_{(\omega_0)}(\omega - \omega_0) + \frac{1}{2}\ddot{\tilde{\varphi}}_{(\omega_0)}(\omega - \omega_0)^2,$$
$$\text{and } \langle \omega - \omega_0 \rangle = 0, \quad (5)$$

where FT means Fourier Transform and differentiation is with respect to $\omega$. Using Parseval's identity, we obtain:

$$\langle t^2 \rangle = \int t^2 |E_{(t)}|^2 dt = \int \tilde{A}^2_{(\omega)} d\omega + \int \dot{\tilde{\varphi}}^2_{(\omega)} \tilde{A}^2_{(\omega)} d\omega, \text{ and}$$
$$\langle t \rangle = \int \dot{\tilde{\varphi}}^2_{(\omega)} \tilde{A}^2_{(\omega)} d\omega \quad (6)$$

Calculating the new pulse duration, the elongation of the pulse due to the added group delay dispersion becomes quite evident [11]:

$$\Delta t^2 = \Delta t^2_{\varphi=0} + \ddot{\tilde{\varphi}}^2_{(\omega_0)} \Delta \omega^2$$
$$= \tau_0^2 + \ddot{\tilde{\varphi}}^2_{(\omega_0)} \frac{(M^2)^2}{4\tau_0^2} \quad (7)$$

Thus, the dispersed pulse form has a larger time-bandwidth product, which we characterize with a dispersion parameter $\kappa^2$, such that:

$$\Delta t^2 \Delta \omega^2 = \frac{(M^2)^2}{4} \kappa^2, \text{ with } \kappa^2 = 1 + \frac{\ddot{\tilde{\varphi}}^2_{(\omega_0)}(M^2)^2}{4\Delta t^4_{\varphi=0}} \quad (8)$$

The characterization of the pulse form $M^2$, dispersion parameter $\kappa^2$, and minimal pulse duration $\tau_0 = \Delta t_{\varphi=0}$ proves to be sufficient to describe and model many of the interesting phenomena associated with femtosecond pulses. For example, propagation of a pulse through dispersive media adds a quadratic phase encompassed in the parameter $\kappa$.

These statistical parameters characterize the pulse during propagation through a dispersive medium in a manner analogous to that commonly employed in the study of beam propagation across different planes. Instead of a CCD camera, as in spatial characterization, two-photon absorption in a GaAsP diode is used.

### III. TWO-PHOTON DIODE RESPONSE

Assuming no spatio-temporal coupling, the photocurrent signal, $S_{NL}$, induced in a two-photon diode is:

$$S_{NL} = \frac{2\beta_{(\omega_0)}}{c\varepsilon_0} \alpha_{\text{spatial}}\, \alpha_{\text{temporal}} \frac{P_m^2 T}{\tau \Delta r^2} \quad (9)$$

$\beta_{(\omega_0)}$ is the transfer function of the diode and is provided by the manufacturer or can be measured. It remains constant assuming that the pulse is long enough that the response is the same for its entire spectrum. $T$ is the period of repetition of the laser and $P_m$ is the average power, easily measured by a linear diode in parallel with the nonlinear diodes. $\tau$ is the duration of the pulse, and $\Delta r$ is the beam waist size at the diode surface, defined in a similar fashion to $\Delta t$ in the spirit of the statistical description. $\alpha_{\text{spatial}}$ and $\alpha_{\text{temporal}}$ are coefficients depending on the form of the pulse.

We experimentally checked the inverse dependence of $S_{NL}$ on the rms definition of pulse duration. After broadening with a stretcher, $\tau$ was calculated both from the measured photocurrent (9) induced by the focused laser beam and from the introduced dispersion by (7). Fig. 1 shows correlation up to a few fs. The accuracy is limited essentially by power fluctuations of the laser and beam size evolution during propagation through the dispersion line.

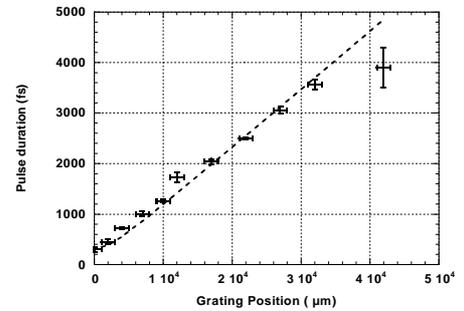

Fig. 1. Pulse duration versus grating distance in a dispersion line. Theoretical results from equation (7) (dashed line), and experimental results (cross) with errors bars. The extreme value suffers from low pulse intensity.

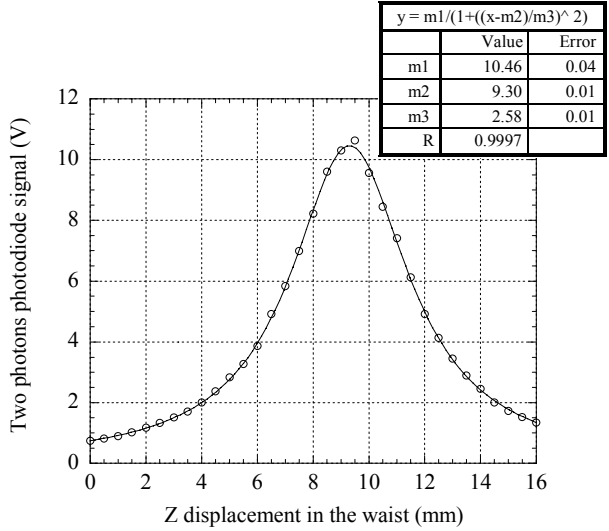

Fig. 2. Nonlinear diode signal versus displacement from the beam waist. Spot size is much smaller than the 2-mm sensitive area. Circles: experimental measurement; line: theoretical fit assuming gaussian laser propagation (10).

Fig. 2 shows the dependence of $S_{NL}$ on the photodiode displacement from the beam waist along the laser's axis of propagation. It agrees precisely with the theoretical beam size during evolution of the gaussian oscillator pulses:

$$\Delta r^2 = w_0^2 \left[ 1 + \left( \frac{z - z_0}{z_R} \right)^2 \right], \qquad (10)$$

where $r_0$ is the beam waist size at the origin $z_0$ and $z_R$ is the Rayleigh range $\pi w_0^2 / \lambda$. From the fit, we deduce a beam waist $w_0 = 25.6$ μm.

## IV. EXTRACTION OF PULSE PARAMETERS FROM TWO-PHOTON CURRENTS

The standard nonlinear diode approach to pulse duration relies on calibration of the signal $S_{NL}$ via an autocorrelation to obtain the transfer function of the diode. This method presupposes interpulse consistency in all parameters, such as the mean power and spot size on the detector, and is thus complicated by the unavoidable fluctuations of all these parameters in time. While it is possible to extract the statistical description of the pulse from an autocorrelation and spectral measurement [11], we propose a simpler experimental setup.

The solution relies on the responses of three two-photon diodes under different added phases in a static experiment. Indeed, without any assumption of form, it is possible to determine the group delay dispersion in a single pulse. Let $S_0$ be the photocurrent signal of the pulse with only the original quadratic phase $A_{chirp}$ present upon exiting the laser aperture. Let $S_1$ and $S_2$ be the signal after the pulse has traversed two known thicknesses $e_1$ and $e_2$ of crystal to add quadratic phases $b_1$ and $b_2$. Unlike an SHG and linear diode autocorrelation, the crystal in our setup does not distort the pulse being measured through any spectral filtering resulting from phase matching requirements. The resolution of the resulting system of equations:

$$S_0^2 = \frac{\alpha}{\Delta t_{\varphi=0}^2 + A_{chirp}^2 \frac{M^4}{4\Delta t_{\varphi=0}^2}} \qquad (11)$$

$$S_1^2 = \frac{\alpha}{\Delta t_{\varphi=0}^2 + (A_{chirp} + b_1)^2 \frac{M^4}{4\Delta t_{\varphi=0}^2}} \qquad (12)$$

$$S_2^2 = \frac{\alpha}{\Delta t_{\varphi=0}^2 + (A_{chirp} + b_2)^2 \frac{M^4}{4\Delta t_{\varphi=0}^2}}, \qquad (13)$$

where $\alpha$ encompasses all of the signal-affecting parameters from (9) that are constant from diode to diode (or, if not constant, whose ratio can be extracted to appropriately weight $S_i$ from the setup without $b_1$, $b_2$), gives us the original quadratic phase $A_{chirp}$:

$$A_{chirp} = \frac{1}{2} \frac{b_2^2 (1/S_1^2 - 1/S_0^2) - b_1^2 (1/S_2^2 - 1/S_0^2)}{b_1 (1/S_2^2 - 1/S_0^2) - b_2 (1/S_1^2 - 1/S_0^2)} \qquad (14)$$

It is noteworthy that by exactly compensating for this group delay dispersion by adding negative phase with a prism configuration, we can obtain the minimal possible pulse duration for this temporal form, $\Delta t_{\varphi=0}^2$. Additionally, we extract from the system the useful relation of pulse duration and form:

$$\tau = \frac{\Delta \omega}{S_0} \left( \frac{b_1 b_2 (b_2 - b_1)}{b_1 (1/S_2^2 - 1/S_0^2) - b_2 (1/S_1^2 - 1/S_0^2)} \right)^{\frac{1}{2}}$$

$$M^2 = 2\Delta\omega (\tau^2 - A_{chirp}^2 \Delta\omega^2)^{\frac{1}{2}} \qquad (15)$$

A spectrogram has to be used in parallel with the three diode configuration to measure the spectral width $\Delta\omega$ of the original pulse. (14) and (15) are established for the same mean power on the three diodes. An experimental measurement is required to calibrate the set-up by using a linear diode to recover the transmission power of beamsplitters and weight $S_0$, $S_1$ and $S_2$ appropriately. Moreover, if they are not precisely known, in order to evaluate precisely the dispersion $b_1$ and $b_2$ (which include the dispersion of the glass pieces, the beamsplitter, and the lenses), we can use the spectral interferometry technique [13], [14] that relies upon the spectrogram of two optical pulses.

## V. CHOICE OF DISPERSIVE MEDIA IN THE SETUP

The choice of dispersive media (dispersion, thickness) is crucial for measuring pulse parameters. If we suppose that the relative error of each signal $S_0$, $S_1$, and $S_2$ is equal to 1%, we can establish the features of the dispersive media. Here the GaAsP photodetector perfectly matches the requirements.

For example, in the case of a gaussian pulse ($M^2 = 1$) that would be 100-fs, were it not lengthened by group delay dispersion, the duration is measured with a relative error of 2% for SF59 glass (dispersion 0.2932 fs²/µm for λ=0.8 µm) with $e_1 = 4$ cm and $e_2 = 13$ cm. For this set of glasses, we are able to measure an initial group delay dispersion between 6000 fs² and 27000 fs² with a relative error less than 20%. The relative error of $M^2$ is approximately 1.8%.

In the case of a 10-fs pulse, another set of glasses must be chosen. For example, with a BK7 glass (dispersion 0.04459 fs²/µm for λ=0.8 µm), we obtain a relative error of pulse duration equal to 2% with $e_1= 3$ mm and $e_2=9$ mm. For this set, the group delay dispersion can be measured between 50 fs² and 300 fs² with a relative error less than 20%, and the accuracy of the $M^2$ factor is unchanged.

The best choice of glass set is dictated by the pulse features, and for improved accuracy, the pulse should be measured with a few sets. The spectral width of the pulse already gives an indication of the glass of choice. This technique is especially well adapted for systems (e.g. amplified lasers) that profit from a minimization of pulse group delay dispersion.

## VI. Measurement of Temporal Width of Non-dispersed Pulse with Known Form Using Two GaAsP Photodiodes

If one makes the usual assumptions for an autocorrelation measurement of no group delay dispersion and known pulse form, laser pulse temporal width can be extracted in a single shot from just two photodiodes in a particularly simple setup. We characterized a Ti:Sapphire oscillator using two GaAsP photodiodes (G1118 Hamamatsu, well adapted for the range 700 to 1100 nm). As shown in the experimental setup (Fig. 3), the laser beam is split in two unbalanced arms, with a high dispersive piece of 5mm ZnSe slab (dispersion ZnSe = 1.03549 fs²/µm) in one arm. Both beams are carefully focused onto the photodiodes.

The signal for photodiode $i \in \{1, 2\}$ is:

$$S_{NL}^{(i)} = \eta_{(\omega_0)}^{(i)} \frac{P_m^{(i)2} T}{\tau_i \Delta r_i^2}, \text{ where}$$

$$\eta_{(\omega_0)}^{(i)} = \frac{2\beta_{(\omega_0)}^{(i)}}{c\varepsilon_0} \alpha_{(spatial)}^{(i)} \alpha_{(temporal)}^{(i)} \quad (16)$$

Photodiode 2 (arm with ZnSe slab) has a broadened pulse $\tau_2^2 = \tau_1^2 + \frac{b^2(M^2)^2}{2\tau_1^2}$, where $b$ is the known dispersion introduced by the ZnSe.

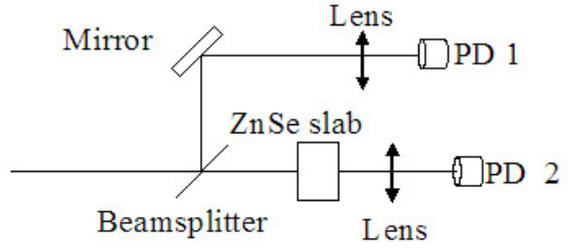

Figure 3. Pulse characterization set-up using two two-photon diodes. S1, S2 are the respective signals of each photodiode PD 1 and PD 2. Lenses are identical in order to avoid spatio-temporal couplings.

The squared ratio between the two photodiodes' signals can be written as:

$$\Gamma_b = \left(\frac{S_{NL}^{(2)}}{S_{NL}^{(1)}}\right)^2$$

$$= \left(\frac{P_m^{(2)} \Delta r_1}{P_m^{(1)} \Delta r_2}\right)^4 \left(\frac{\eta_{(\omega_0)}^{(2)}}{\eta_{(\omega_0)}^{(1)}}\right)^2 \frac{\tau_1^2}{\frac{b^2(M^2)^2}{2\tau_1^2} + \tau_1^2} \quad (17)$$

$$= \Gamma_0 \frac{\tau_1^2}{\frac{b^2(M^2)^2}{2\tau_1^2} + \tau_1^2}$$

The calibration coefficient ratio $\Gamma_0$ is determined by repetitive measurements at the working wavelength (800 nm) of the diode in the setup without the ZnSe slab (b=0). From the measurements of the ratio $\Gamma_b$ with the 5-mm ZnSe slab, we can then deduce the temporal width of the laser:

$$\tau_1^2 = \frac{bM^2 \Gamma_b^{\frac{1}{2}}}{[2(\Gamma_0 - \Gamma_b)]^{\frac{1}{2}}} \quad (18)$$

Assuming a gaussian shape ($M^2=1$), the FWHM temporal width $\tau_{FWHM}$ is deduced from $\tau_1$ using the relation $\tau_{FWHM}^2 = \tau_1^2 16\ln 2$ for a gaussian pulse. In our experiment, $\Gamma_b = 2.5$ and $\Gamma_0 = 10.3 \pm 0.1$, yielding the calculated $\tau_{FWHM}$ of 115 fs. This agrees well with the autocorrelation measurement of $\tau_{FWHM}$ as 120 fs. The experimental measurement range is here limited to 50-150 fs (Fig. 4). The lower limit arises from a low photocurrent in diode 2, while the upper limit is due to limited broadening causing an inaccurate ratio.

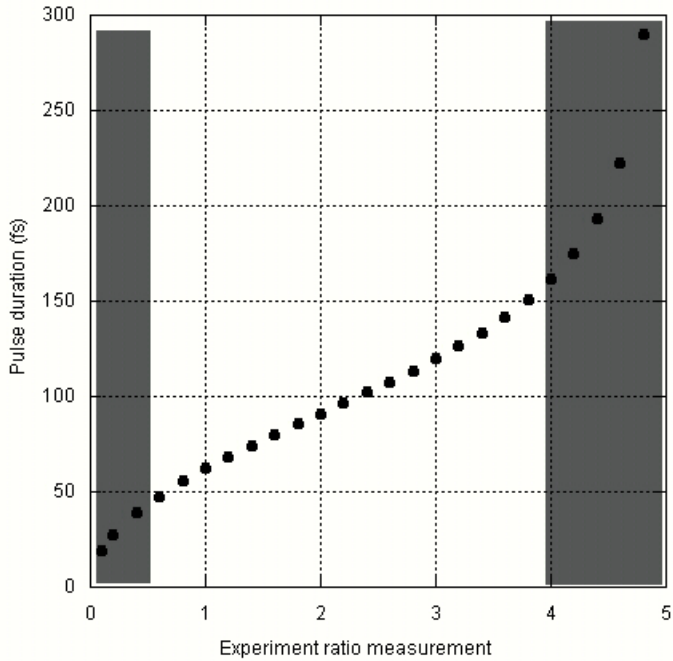

Figure 4. Pulse width versus ratio measurement ($\Gamma_b$) in our experimental conditions. Gray area: unavailable zone for the setup due to low SNR.

Fig. 5 shows several measurements of the same laser for different input power in the setup. Sensitivity limits are due to either a very small photocurrent (due to dark current and thermal noise) or amplifier saturation. Measurement of 1mW average power is possible with this setup.

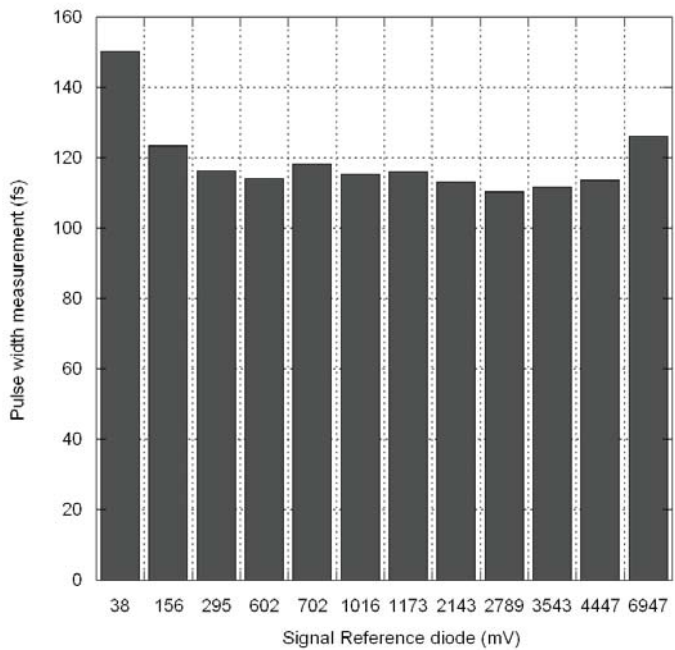

Fig. 5. Measurement of TiSapphire laser pulse duration for various input powers.

## VII. CONCLUSION

A statistical description proves to be useful in describing femtosecond laser pulse evolution through dispersive media. Further, a complete characterization of the pulse in terms of form, group delay dispersion, and duration is achievable in a simple static experimental setup with three nonlinear diodes. No assumptions are required to extract group delay dispersion information, and if a spectrogram is available, form and duration can also be measured without the conventional assumptions. Using only two diodes as described here, we can extract pulse duration for a non-dispersed, gaussian pulse in a single shot, and results are in excellent agreement with autocorrelation measurements.